\documentclass{PoS}

\title{Light charged Higgs in the NMSSM}

\ShortTitle{Light charged Higgs in the NMSSM}

\author{Radovan DERMISEK\thanks{Based on work done in collaboration with J. Gunion, E. Lunghi and A. Raval.}\\
       Physics Department, Indiana University, Bloomington, IN 47405, USA\\
       E-mail: \email{dermisek@indiana.edu}}

\abstract{We review motivation for a light charged Higgs scenario that occurs in a class of models with a light CP-odd Higgs boson. If the light CP-odd Higgs is doublet-like the charged Higgs is typically lighter than the top quark and it dominantly decays into $W^\pm A$ with  $B(H^+ \to W^+ A) > 70 \%$.
We summarize experimental constraints from direct searches and discuss search strategies based on subleading decay modes of the CP-odd Higgs. 
A search for $t \to H^+ b, \;\;\;\; H^+ \to W^+ A, \;\;\;\; A \to \mu^+ \mu^- $
 could provide an evidence for the charged Higgs or a discovery already with 1 fb$^{-1}$ of data at the LHC.
 }

\FullConference{Prospects for Charged Higgs Discovery at Colliders \\
                27-30 September 2010\\
                Uppsala University, Sweden}

\begin{document}

\section{Motivation for the light charged Higgs scenario}

A light charged Higgs appearing in top quark decays, $t \to H^+ b$,  that dominantly decays into $W^+$ boson and   the CP-odd Higgs boson $A$,  occurs in a class of models that were suggested to 
eliminate the fine tuning problem in supersymmetric models associated with non-observation of the standard-model-like  Higgs boson at LEP.

The LEP exclusion limits, $m_h > 114$ GeV, constraints from precision electroweak data, $m_h < 157$ GeV~\cite{lepewwg}, and recently also the Tevatron limits,  leave about 40 GeV window for the standard model (SM) Higgs boson.
This 40 GeV window is very interesting and there are several suggestive hints or coincidences related to it. This window overlaps with the range of  Higgs masses in which the standard model  can be a consistent theory all the way to the grand unification scale or the Planck scale, $m_h \simeq 125 - 175$ GeV. 
This window also overlaps with the range of Higgs masses predicted in the minimal supersymmetric model (MSSM), $m_h \lesssim 135$ GeV. 
 Hence, it is expected that the  Higgs boson is somewhere in this window and most of the effort is focused on discovery strategies related to this possibility. 
 
 However there are also several compelling hints that the SM-like Higgs boson is below the LEP exclusion  limits. First of all, electroweak symmetry breaking (EWSB) in  simple supersymmetric (SUSY) extensions of the standard model, with superpartners near the electroweak (EW) scale, generically predicts the Higgs boson  not heavier than about 100 GeV. Non-observation of the Higgs boson at LEP resulted in the ``fine-tuning" problem in these models~\cite{Dermisek:2009si}. Second of all, the best fit to  precision electroweak data is achieved for the Higgs mass of 87 GeV~\cite{lepewwg}. The third hint comes from the LEP data: the largest excess of Higgs like events at LEP corresponds to the Higgs mass of 98 GeV.
It is an interesting coincidence that natural EWSB in SUSY models, precision EW data, and the largest excess of Higgs like events point to the same region. This  supports the idea that  the Higgs boson is light, somewhere near 100 GeV, and we missed it at LEP. How can this be?

The basic idea is very simple: if the SM-like Higgs boson decays in a different way than the Higgs boson in the standard model then the usual experimental limits do not apply. Such a Higgs can be light, as predicted from SUSY, gives better agreement with precision electroweak data, and  can even explain the largest excess of Higgs like events at LEP~\cite{Dermisek:2005ar,Dermisek:2005gg,Dermisek:2007yt}. 

In theories beyond the SM the Higgs sector is usually  more complicated and there are typically many other Higgses in addition to the SM-like Higgs boson. For example, there are five Higgs bosons in the MSSM:  light and heavy CP-even Higgses, $h$ and $H$, the CP-odd Higgs, $A$, and a pair of charged Higgs bosons, $H^\pm$; seven in the next-to-minimal supersymmetric model (NMSSM),  which contains an additional singlet field: three CP-even Higgs bosons, $h_{1,2,3}$, two CP-odd Higgs bosons, $a_{1,2}$, and a pair of charged Higgs bosons; and there are many simple models with even more complicated Higgs sectors.
Usually we explore parameter space in which the extra Higgses are somewhat heavy -- the so called decoupling limit. 

The decoupling limit is not  the only possibility.
One of the extra Higgses can be light, for example, the singlet CP-odd Higgs in the NMSSM. If it is sufficiently light, the SM-like Higgs boson can (and typically would) dominantly decay into a pair of CP-odd Higgses and eventually, depending on the mass of the CP-odd Higgs boson, into four b quarks, four $\tau$ leptons,  four c quarks,  four $\mu$ leptons,  four electrons,  four light  quarks, or gluons~\cite{Dermisek:2005ar}. Limits on these 4-body final states are weaker than limits on the SM Higgs boson (decaying into $b \bar b$) and currently $4 \tau$ leptons, c quarks,  and light  quarks or gluons  final states allow the SM-like Higgs boson at $\sim 100$ GeV or even lighter~\cite{Chang:2008cw,Dermisek:2010mg}, as is predicted from the best theoretically motivated region of the parameter space in supersymmetric theories, and it also gives much better agreement with precision electroweak data. In addition, the subleading decay mode of the Higgs boson, $h \to b \bar b$, with branching ratio of $\sim 10 \%$ can completely explain the 
largest excess ($2.3 \sigma$) of Higgs-like events at LEP in the
$b\bar b$ final state (for a reconstructed mass $M_{b\bar b}\sim 98$ GeV)~\cite{Dermisek:2005gg}.

Another, perhaps even more interesting 
variation of the above NMSSM scenario is the scenario with a {\bf doublet-like} CP odd Higgs bellow the $b \bar b$ threshold.
For small  $\tan \beta$, $\tan \beta \lesssim 2.5$, this scenario is  the least constrained (and only  marginally ruled out) in the MSSM, and thus easily viable in simple extensions of the MSSM~\cite{Dermisek:2008id,Dermisek:2008sd,Dermisek:2008uu,Bae:2010cd}. Surprisingly, the  prediction from this region is that
all the Higgses resulting from two Higgs doublets: $h$, $H$, $A$ and $H^\pm$ could have been produced already at LEP or the Tevatron, but would have escaped detection because they decay in modes that have not been searched for or the experiments are not sensitive to.
The heavy CP even and the CP odd Higgses could have been produced at LEP in $e^+ e^- \to H A$ but they would avoid detection because the dominant decay mode of $H$,  $H \to ZA$,  has not been searched for. 
The charged Higgs is also very little constrained although it could have been pair produced at LEP or appeared in decays of top quarks produced at the Tevatron. 
{\bf The dominant decay mode of the charged Higgs in this scenario is $H^\pm \to W^{\pm} A$ with $A \to \tau^+ \tau^-$ or $A \to c \bar c$}.
 In addition, the charged Higgs with  properties emerging in this scenario and the mass close to the mass of the $W$ boson
could explain the $2.8 \sigma$ deviation from lepton universality in $W$ decays measured at LEP~\cite{:2004qh} as pointed out  in Ref.~\cite{Dermisek:2008dq}.

The main idea is simple, and it can be realized in a variety of other models. For example, in specific little Higgs models the SM-like Higgs boson can dominantly decay into four c quarks~\cite{Bellazzini:2009kw} or four gluons~\cite{Bellazzini:2009xt}. Four body final states of the Higgs boson can also occur in composite Higgs models~\cite{Gripaios:2009pe}, and models for dark matter~\cite{Mardon:2009gw} among others.  For a review of other scenarios and references, see Ref.~\cite{Chang:2008cw}. Additional level of complexity arises when the Higgs (extra) sector is more complicated, see {\it e.g.} Ref.~\cite{Dermisek:2006wr}, or when several Higgs bosons share the coupling to the Z boson and there is not a single SM-like Higgs boson, see {\it e.g.}~\cite{Dermisek:2007ah}.

Unusual decay modes require unusual search strategies. Examples of search strategies for the SM-like Higgs and the light CP-odd Higgs in the above scenarios were discussed in Refs.~\cite{Lisanti:2009uy,Dermisek:2006py,Dermisek:2009fd,Dermisek:2010tg}.
In this talk we will focus on the charged Higgs. Let's start with a brief summary of basic features of the charge Higgs sector.

\section{Basic features of the light charged Higgs scenario in the (N)MSSM and beyond}

\begin{figure} 
\vspace{-1.cm}
\begin{center}
\includegraphics[width=.55\textwidth]{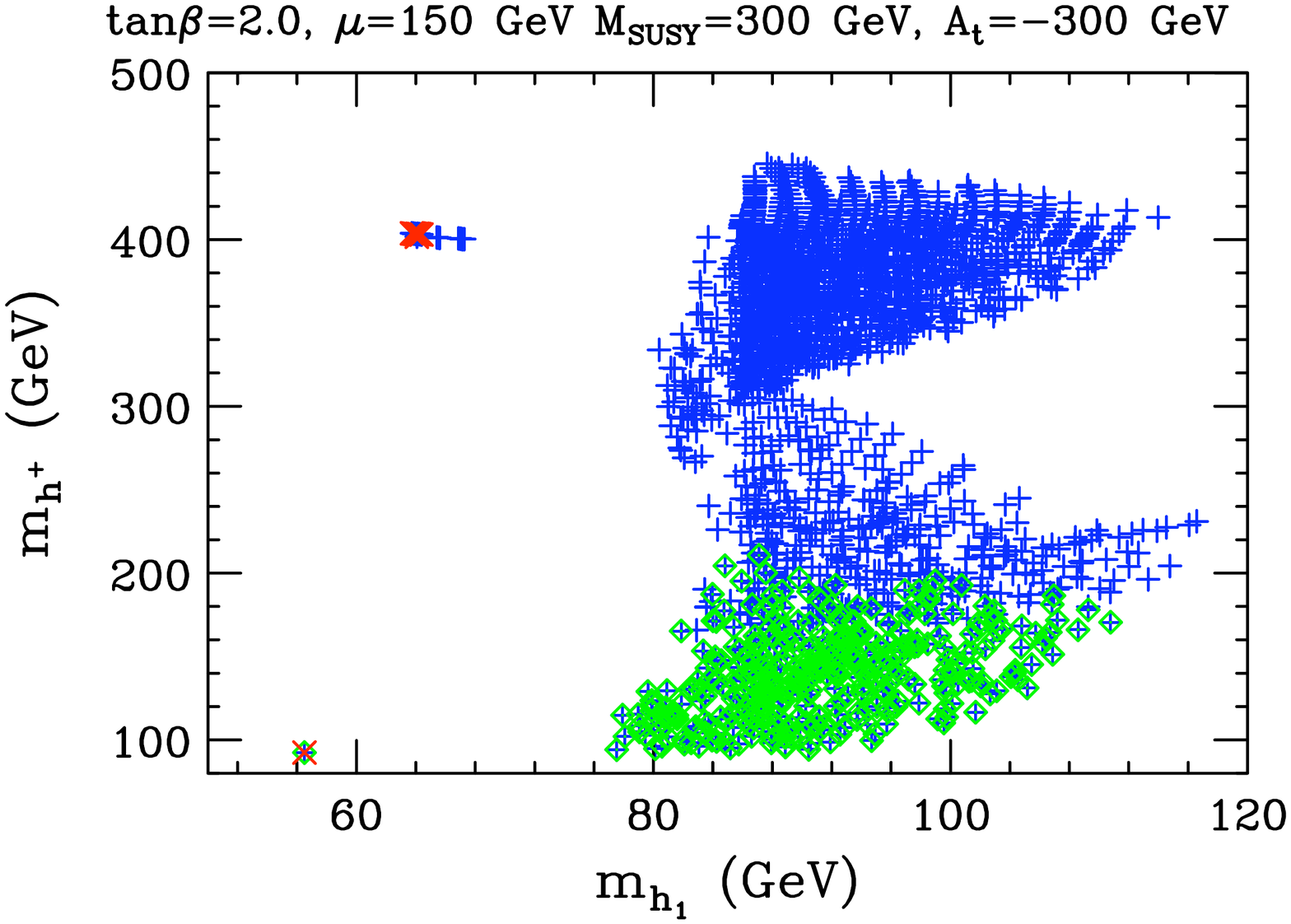}
\hspace{-1.7cm}
\includegraphics[width=.55\textwidth]{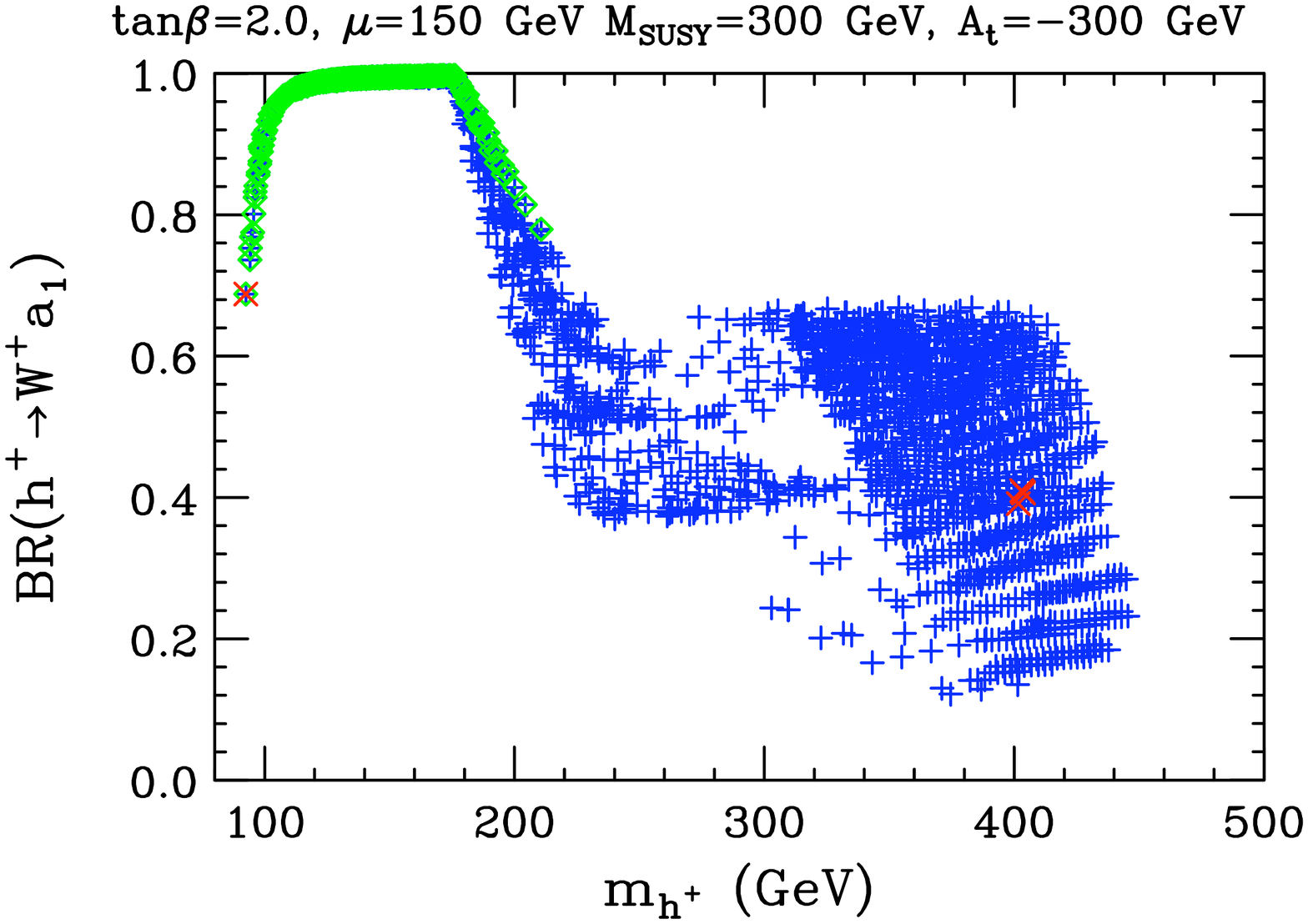} 
\end{center}
\vspace{-.5cm}
\caption{The mass of the charged Higgs  vs. the mass of the lightest CP-even Higgs boson (left) and $B(H^+ \to W^+ a_1)$ vs. the mass of the charged Higgs (right) in the NMSSM for parameters specified in the plots. These plots are from Ref.~\cite{Dermisek:2008uu}. Green points represent scenarios with a doublet-like CP-odd Higgs.} 
\label{fig1} 
\end{figure}

In the MSSM the mass of the charged Higgs is related to the mass of the $W$ and the mass of the CP-odd Higgs:
\begin{equation}
m^2_{H^\pm} = m^2_{W^\pm} + m_A^2 + ({\rm SUSY \; loops}). 
\end{equation}
The SUSY contributions to this relation are not significant and thus, for small $m_A$, we find $m_{H^\pm} \simeq m_{W^\pm}$. The decay mode $H^\pm \to W^{\pm \star} A$ (with off-shell $W$) still dominates, $B(H^+ \to W^+ A) \gtrsim 70 \%$. The subleading decay mode of the charged Higgs  is $\tau \nu$ or $cs$ depending on $\tan \beta$. This charged Higgs boson scenario is not ruled out even in the MSSM by direct searches for the charged Higgs. It is ruled out however (only barely) by searches for the SM-like Higgs which is too light in this region of the parameter space~\cite{Dermisek:2008id,Dermisek:2008sd}.\footnote{A similar scenario with a light charged Higgs  that dominantly decays into $ W^{\pm \star} A$ in the MSSM with $\tan \beta < 1$ was  discussed in Ref.~\cite{Akeroyd:2002hh}.}

In models beyond the MSSM the tight mass relations between Higgs bosons are relaxed and the scenario with a light charged Higgs dominantly decaying into $W^{\pm} A$ can be viable~\cite{Dermisek:2008id,Dermisek:2008sd}. For example, in the NMSSM the CP-even Higgs mass gets additional contribution that can push it above the LEP limits, or the Higgs can have additional decay modes allowing it to avoid  LEP limits.
In addition, there are two CP-odd Higgses in the spectrum and the lightest mass eigenstate is a linear combination of the doublet CP-odd Higgs and the  singlet CP-odd Higgs. When the light CP-odd Higgs has a significant doublet component the scenario resembles the MSSM scenario discussed above (and it can be phenomenologically viable), see Fig.~\ref{fig1} (green points represent scenarios with a doublet-like CP-odd Higgs). Increasing the singlet component in the light CP-odd Higgs the $B(H^+ \to W^+ A)$ is decreasing and the charged Higgs resembles more and more the charged Higgs in the decoupling limit of the MSSM with  the usual $\tau \nu$ or $cs$ decay modes. For a detailed analysis of this scenario in the NMSSM see Ref.~\cite{Dermisek:2008uu}. This scenario can also be viable in  general beyond the MSSM scenarios with two Higgs doublets and new physics parameterized by higher-dimensional operators~\cite{Bae:2010cd}.

\section{Constraints and prospects at the LHC with 1 fb$^{-1}$}

\begin{figure} 
\begin{center}
\includegraphics[width=.4\textwidth]{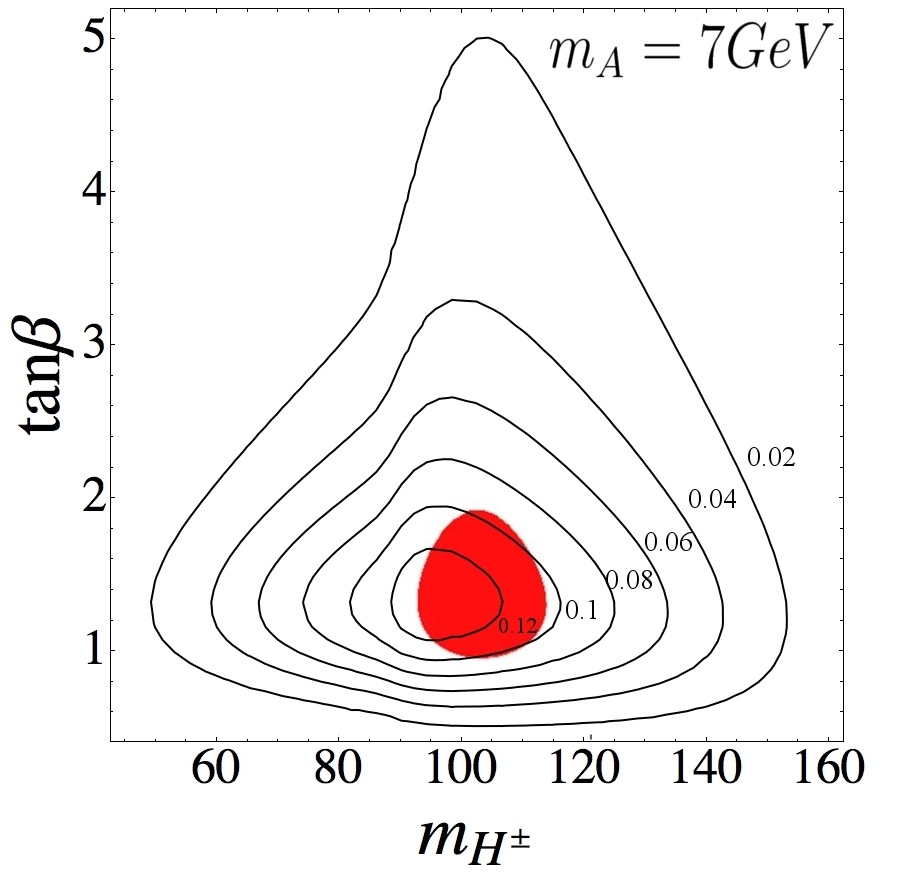} 
\end{center}
\caption{Contours of constant $B( t \to H^+ b) \times B(H^+ \to W^+ A) \times B( A\to \tau^+ \tau^-)$ in $m_{H^+} - \tan \beta$ plane with all  couplings as in the MSSM with the CP odd Higgs mass $m_A = 7$ GeV. The shaded region is excluded by the CDF search~\cite{CDF_charged_higgs}. This figure is from Ref.~\cite{DLR}.
} 
\label{fig2} 
\end{figure}


If the CP-odd Higgs boson has a significant dublet component than the charged Higgs is generically  light, typically lighter  than the top quark.
Depending on the  mass of the charged Higgs and $\tan \beta$ the  $B(t \to H^+ b)$  can go up to 40\% (for $\tan \beta =1$ and $m_{H^\pm} = 80$ GeV) dropping very fast with increasing  $\tan \beta$ and increasing the mass of the charged Higgs. 
 The dominant decay mode of the charged  Higgs  is $H^+ \to W^+ A$ which is typically close to 100\% except when the mass of the charged Higgs gets close to the mass of $W^+$~\cite{Dermisek:2008id,Dermisek:2008uu,Bae:2010cd}. Prediction for $B( t \to H^+ b) \times B(H^+ \to W^+ A) \times B( A\to \tau^+ \tau^-)$ 
 as a function of $\tan \beta$ and $m_{H^+}$ assuming all couplings as in the MSSM for $m_A = 7$ GeV is given in Fig.~\ref{fig2}. 
 
 The CDF recently performed a search and set the upper limit on $B( t \to H^+ b) \times B(H^+ \to W^+ A) \times B( A\to \tau^+ \tau^-)$ to about 10\%~\cite{CDF_charged_higgs} excluding a small region of the parameter space. The exact impact of the CDF search is also plotted in Fig.~\ref{fig2} (the shaded region is excluded).
 
 
 The search for the charged Higgs decaying into $W^\pm A$ will be relatively easy  at the LHC  which is a top quark factory. It is  advantageous to look for a subleading decay mode of the CP-odd Higgs, $A \to \mu^+ \mu^-$. In addition, we can look for events in which one of the $W$ bosons decays into $\mu \nu$, resulting in 3-muon events with properties that easily stand out from the background.  
  For a reference point in the parameter space specified by $m_{H^+} = 100$ GeV, $\tan \beta = 2$ and $m_A = 7$ GeV with 1 fb$^{-1}$ of data one expects about 30 signal 3-muon events with the invariant mass of the two closest muons  (given by the mass of the CP odd Higgs)  in one 100 MeV bin (the resolution of the  di-muon invariant mass).  In order to suppress the background dominated by semi-leptonic bottom and charm decays and Drell-Yan processes we  impose cuts on $p_T$ of muons, $p_T> 10$ GeV.
 This leaves about 5 signal events in one 100 MeV bin of di-muon mass with total background of about 0.5 events per 100 MeV~\cite{DLR}.  This analysis can be further improved by adding the signal events  in which one of the $W^\pm$ decays into $e^\pm$ instead of $\mu^\pm$. In addition, the background can be further suppressed to a negligible level by requiring at least one b in events. 
  
\vspace{.3cm}

In conclusion, in scenarios with a light doublet-like CP-odd Higgs boson the charged Higgs is typically lighter than the top quark and it dominantly decays into $W^\pm A$ with  $B(H^+ \to W^+ A) > 70 \%$. Searching for  dominant decay modes of $A$ typically requires several tens of fb$^{-1}$ of data. However, searching for subleading decay modes is very promising with early data at the LHC. Especially the search for 
$t \to H^+ b, \;\;\;\; H^+ \to W^+ A, \;\;\;\; A \to \mu^+ \mu^- $
 could provide an evidence for the charged Higgs or  a discovery already with 1 fb$^{-1}$ of data.

\section*{Acknowledgments}
I would like to thank J. Gunion, E. Lunghi and A. Raval for collaboration on projects this talk is based on.


\begin{thebibliography}{99}
 
 
  \bibitem{lepewwg}
  LEP-EWWG, http://lepewwg.web.cern.ch/LEPEWWG.

  
\bibitem{Dermisek:2009si}
  For a review and references, see {\it e.g.} R.~Dermisek,
  Mod.\ Phys.\ Lett.\  A {\bf 24}, 1631 (2009)
  [arXiv:0907.0297 [hep-ph]].


\bibitem{Dermisek:2005ar}
  R.~Dermisek and J.~F.~Gunion,
  Phys.\ Rev.\ Lett.\  {\bf 95}, 041801 (2005)
  [arXiv:hep-ph/0502105].
  
\bibitem{Dermisek:2005gg}
  R.~Dermisek and J.~F.~Gunion,
  Phys.\ Rev.\  D {\bf 73}, 111701 (2006)
  [arXiv:hep-ph/0510322].
  
  
\bibitem{Dermisek:2007yt}
  R.~Dermisek and J.~F.~Gunion,
  Phys.\ Rev.\  D {\bf 76}, 095006 (2007)
  [arXiv:0705.4387 [hep-ph]].
  
  
 
\bibitem{Chang:2008cw}
  S.~Chang, R.~Dermisek, J.~F.~Gunion and N.~Weiner,
  Ann.\ Rev.\ Nucl.\ Part.\ Sci.\  {\bf 58}, 75 (2008)
  [arXiv:0801.4554 [hep-ph]].

  
\bibitem{Dermisek:2010mg}
  R.~Dermisek and J.~F.~Gunion,
  Phys.\ Rev.\  D {\bf 81}, 075003 (2010)
  [arXiv:1002.1971 [hep-ph]].
  
  
 
  
\bibitem{Dermisek:2008id}
  R.~Dermisek,
  arXiv:0806.0847 [hep-ph].

\bibitem{Dermisek:2008sd}
  R.~Dermisek,
  AIP Conf.\ Proc.\  {\bf 1078}, 226 (2009)
  [arXiv:0809.3545 [hep-ph]].
  
 
  
  
\bibitem{Dermisek:2008uu}
  R.~Dermisek and J.~F.~Gunion,
  Phys.\ Rev.\  D {\bf 79}, 055014 (2009)
  [arXiv:0811.3537 [hep-ph]].
 

 
\bibitem{Bae:2010cd}
  K.~J.~Bae, R.~Dermisek, D.~Kim, H.~D.~Kim and J.~H.~Kim,
  arXiv:1001.0623 [hep-ph].

  
\bibitem{:2004qh}
    [LEP Collaborations],
  arXiv:hep-ex/0412015.

  
\bibitem{Dermisek:2008dq}
  R.~Dermisek,
  arXiv:0807.2135 [hep-ph].
  
  
  
  
\bibitem{Bellazzini:2009kw}
  B.~Bellazzini, C.~Csaki, A.~Falkowski and A.~Weiler,
  Phys.\ Rev.\  D {\bf 81}, 075017 (2010)
  [arXiv:0910.3210 [hep-ph]].


\bibitem{Bellazzini:2009xt}
  B.~Bellazzini, C.~Csaki, A.~Falkowski and A.~Weiler,
  arXiv:0906.3026 [hep-ph].
  
\bibitem{Gripaios:2009pe}
  B.~Gripaios, A.~Pomarol, F.~Riva and J.~Serra,
  JHEP {\bf 0904}, 070 (2009)
  [arXiv:0902.1483 [hep-ph]].

\bibitem{Mardon:2009gw}
  J.~Mardon, Y.~Nomura and J.~Thaler,
  arXiv:0905.3749 [hep-ph].


  
  
\bibitem{Dermisek:2006wr}
  R.~Dermisek and J.~F.~Gunion,
  Phys.\ Rev.\  D {\bf 75}, 075019 (2007)
  [arXiv:hep-ph/0611142].


   
\bibitem{Dermisek:2007ah}
  R.~Dermisek and J.~F.~Gunion,
  Phys.\ Rev.\  D {\bf 77}, 015013 (2008)
  [arXiv:0709.2269 [hep-ph]].



\bibitem{Lisanti:2009uy}
  M.~Lisanti and J.~G.~Wacker,
  Phys.\ Rev.\  D {\bf 79}, 115006 (2009)
  [arXiv:0903.1377 [hep-ph]].

  
\bibitem{Dermisek:2006py}
  R.~Dermisek, J.~F.~Gunion and B.~McElrath,
  Phys.\ Rev.\  D {\bf 76}, 051105 (2007)
  [arXiv:hep-ph/0612031].


\bibitem{Dermisek:2009fd}
  R.~Dermisek and J.~F.~Gunion,
  Phys.\ Rev.\  D {\bf 81}, 055001 (2010)
  [arXiv:0911.2460 [hep-ph]].
  
\bibitem{Dermisek:2010tg}
  for a brief review, see also, R.~Dermisek,
  arXiv:1008.0222 [hep-ph].
  
\bibitem{Akeroyd:2002hh}
  A.~G.~Akeroyd, S.~Baek, G.~C.~Cho and K.~Hagiwara,
  Phys.\ Rev.\  D {\bf 66}, 037702 (2002)
  [arXiv:hep-ph/0205094].

   \bibitem{CDF_charged_higgs}
  R. Erbacher, A. Ivanov, and W. Johnson, CDF, 2010, \\
  http://www-cdf.fnal.gov/physics/new/top/2009/tprop/nMSSMhiggs/.
  
  
  \bibitem{DLR}
R. Dermisek, E. Lunghi and A. Raval, in progress.
  

\end{thebibliography}
\end{document}